\documentclass[11pt]{article}
\usepackage[utf8]{inputenc}
\pdfoutput=1
\usepackage{amssymb,amsmath,mathrsfs,enumerate}
\usepackage{graphicx,rotate,multicol}
\usepackage{float}
\usepackage{tocloft}
\usepackage{subfig}
\usepackage[margin=10pt,labelfont=bf]{caption}
\usepackage{cite}
\usepackage{soul}
\usepackage[colorlinks=true,
            linkcolor=blue,
            urlcolor=blue,
            citecolor=teal]{hyperref}
\usepackage{multirow}

\makeatletter
\let\Hy@linktoc\Hy@linktoc@page
\makeatother

\usepackage{color}
\definecolor{ourcolor}{rgb}{0.7, 0.25, 0.05}
\usepackage{tikz,braket}
\long\def\rpl#1!!#2!!{\textcolor{red}{#1} \textcolor{blue}{#2}}

\def \order(#1){{\mathcal O} \left(#1 \right)}

\evensidemargin=\oddsidemargin
\allowdisplaybreaks

\title{\color{black}{\bf Prospects of Migdal Effect in the Explanation of XENON1T Electron Recoil Excess}}

\author {\bf Ujjal Kumar Dey,$^{a,}$\footnote{ujjal@iiserbpr.ac.in} 
\hspace{4pt}  Tarak Nath Maity,$^{b,}$\footnote{tarak.maity.physics@gmail.com}
\hspace{4pt}  Tirtha Sankar Ray$^{b,c,}$\footnote{tirthasankar.ray@gmail.com} 
\\[10pt]
\small\em $^a$Department of Physical Sciences, Indian Institute of Science Education and Research, Berhampur 760010, India\\
\small\em $^b$Department of Physics, Indian Institute of Technology, Kharagpur 721302, India\\
\small\em $^c$Centre for Theoretical Studies, Indian Institute of Technology, Kharagpur 721302, India
}

\date{}

\begin{document}

\maketitle

\begin{abstract}
The XENON1T experiment has recently announced the observation of an excess in electron recoil events at energy range of $1-7$ keV with a $3.5~\sigma$ signal significance over the Standard Model prediction. In this letter we sketch the prospects of explaining such an excess from Migdal ionization events with below threshold nuclear recoil energies. Interestingly, these are expected to show signal events in the ballpark energy scale of the observed excess. We demonstrate that  the observed signal can be reproduced through the Migdal effect by an $\mathcal{O}(1)$ GeV neutron-philic dark matter having a spin-dependent coupling with the nucleus. A more optimistic scenario is explored where the Migdal ionization is driven by MeV scale boosted dark matter.
\end{abstract}


\newpage

\hrule \hrule
\tableofcontents
\vskip 10pt
\hrule \hrule 

\section{Introduction}
\label{sec:xenon}
The XENON collaboration recently reported the results of new physics searches with low-energy electronic recoil data obtained in the XENON1T detector~\cite{Aprile:2020tmw}. They observed an excess at lower electron recoil energies 1-7 keV mostly crowded toward the lower energy ranges of 2-4 keV range, with 285 observed events as compared to 232$\pm$15 events expected from known backgrounds.  Neglecting a possible new source of contamination from a tritium impurity in the detector the statistical significances of the excess is at  3.5 $\sigma$ when interpreted in terms of solar axions, 3.2 $\sigma$ with enhanced neutrino magnetic moment considerations\cite{Aprile:2020tmw, Bell:2005kz, Bell:2006wi}.  
The significances  decrease dramatically once an unconstrained tritium component is included in the  background analysis.  The other generic concern for the reported excess is the location which is very close to the threshold of the experiment that may distort the performance of the detector in these energy scales. However this result has triggered several proposals to explain the reported anomaly including  scenarios with  hidden dark photon dark matter (DM)~\cite{Alonso-Alvarez:2020cdv, Choi:2020udy}, Co-SIMP DM~\cite{Smirnov:2020zwf}, boosted DM~\cite{Kannike:2020agf, Fornal:2020npv, Chen:2020gcl, Su:2020zny}, non-relativistic DM with Rayleigh operators~\cite{Paz:2020pbc}, axion-like particle dark matter~\cite{Takahashi:2020bpq}, argon decays~\cite{Szydagis:2020isq}, and non-standard neutrino electron interaction~\cite{Boehm:2020ltd, AristizabalSierra:2020edu}.

At the XENON  experiment an efficient discrimination of the nuclear recoil and electronic recoil events are done by using the  signal ratio parameter, S1/S2. It is expected that in nuclear recoil most of the energy  will be dissipated  through prompt photons detected mostly in the PMT located at the bottom of the detector material having a relatively large S1 signal. While events from electron recoil have a larger charge deposition  through ionisation. These  free electron then drift through the material in the presence of external electric field producing a delayed  bremsstrahlung photons detected at the top  PMT detectors making it the major constituent of  the so-called  S2 signals~\cite{Aprile:2020tmw}. An interesting possibility is a scenario where nuclear recoil events lead to considerable ionisation of the Xenon atoms  but the nuclear recoil energies remain below the detector threshold. These Migdal ionization events essentially contributing to the electron recoil signal~\cite{Ibe:2017yqa, Dolan:2017xbu, Bell:2019egg, GrillidiCortona:2020owp}  may offer a possible explanation to the observed excess.

Note that for the $n=3$ and $n=2$ atomic states of the Xenon atom, the ionization energies are given by $0.66$ keV and $4.9$ keV respectively. This strikingly coincides with the generic energy scale of the observed  XENON1T excess around 1-7 keV.   Thus an understanding of the origin  of the  energy scale associated with the excess distinguishes the Migdal  framework.
In this letter, we make a systematic study of a possible explanation of the   observed excess events by considering  the Migdal  ionization effect with sub-threshold nuclear recoil. Remaining agnostic about the specific features of the DM particle and the underlying model thereof,  we explore scenarios where the DM nucleus scattering cross-section is spin-independent and spin-dependent as an explanation to the recently observed XENON1T excess. Additionally we explore the possibility of boosted DM driven Migdal effect.

The rest of the letter is organised as follows.  In Section \ref{sec:migdal} we briefly review the Migdal effect. In the subsections \ref{subsec:normDM} and \ref{subsec:boostDM} we explore the possibility to fit the observed excess utilising the Migdal effect before we conclude.

\section{Migdal effect and the explanation of the excess} 
\label{sec:migdal}
When a DM scatters with the nucleus of the detector atoms there is a time lag for the electron cloud to catch-up the recoiling nucleus, which results in a charge asymmetry. This leads to an ionization and/or de-excitation of the atoms, eventually contributing to the electron recoil signals in direct detection experiments. For light DM where the nuclear recoil is below threshold these ionization (and sub-dominant bremsstrahlung process) signals provide a novel handle to search for them~\cite{Vergados:2004bm, Moustakidis:2005gx, Ejiri:2005aj, Bernabei:2007jz, Vergados:2013raa, Ibe:2017yqa, Dolan:2017xbu, Bell:2019egg, Baxter:2019pnz, Essig:2019xkx}.  This process of ionization or excitation of an atom is known as Migdal effect. In this section we briefly review the methodology of the estimation of the rate of ionization events due to the Migdal effect following Refs.~\cite{Ibe:2017yqa, Dolan:2017xbu, Bell:2019egg} and utilising this framework to present a possible explanation of the recently observed XENON1T excess.
Consider a dark matter particle $\chi$ with mass $m_{\chi}$ collides with the nucleus $N$. The differential event rate for a nuclear recoil of energy $E_R$ followed by an ionisation electron with energy $E_e$ is given by,
\begin{align}
\label{eq:ionrate}
\frac{d^{2}R_{\rm ion}}{dE_{R}dE_{e}} =
	\frac{dR_{\chi N}}{dE_{R}} 
	\left|Z_{\rm ion}\right|^{2}, 
\end{align}
where the first piece is the differential rate for DM-nucleus scattering $R_{\chi N}$ w.r.t. $E_{R}$. This can be given as,
\begin{align}
\label{eq:dmN}
\frac{dR_{\chi N}}{dE_{R}} = \eta(v^{\rm min}_{\chi})\left(\frac{\rho ~\sigma^{\rm SI/SD}_{\chi N}}{2\mu_{N}^{2}m_{\chi}}\right),
\end{align}
where $\rho$ is the local DM density, fixed at $0.3$ $\rm GeV/cm^3$. $\sigma^{\rm SI/SD}_{\chi N}$ is the spin-(in)dependent DM-nucleus scattering cross-section, and the reduced mass of the DM-nucleus system $\mu_{N} = m_{\chi}m_{N}/(m_{\chi}+m_{N})$. $\eta(v^{\rm min}_{\chi})$ is the velocity average of the inverse DM speed. To estimate $\eta(v^{\rm min}_{\chi})$, we have used standard Maxwell-Boltzmann velocity distribution, truncated at the galactic escape velocity ($v_{\rm esc}$). The chosen values $v_{\rm esc}$ and $v_0$ are $544~ \rm km/s$ and $220 ~\rm km/s$ respectively. The expression for $v^{\rm min}_{\chi}$ is given by
\begin{equation}
v^{\rm min}_{\chi}=\sqrt{\frac{m_N E_R}{2 \mu_{N}}}+ \frac{E_e+E_{nl}}{\sqrt{2 m_N E_R}},
\end{equation}
where $E_{nl}$ is the binding energy of each states. The DM-nucleus scattering cross-section $\sigma^{\rm SI/SD}_{\chi N}$ can easily be related to the DM-nucleon normalized cross section~\cite{Lewin:1995rx}. The nuclear form factor is ignored in writing Eq.~\eqref{eq:dmN} since we are interested in small momentum transfers.
The second piece in Eq.~\eqref{eq:ionrate}, $|Z_{\rm ion}|^{2}$ represents the differential ionization rate corresponding to single electron ionization and is given in terms of the ionization probability $p^{c}_{q_{e}}$ as,
\begin{align}
\label{eq:zion}
\left|Z_{\rm ion}\right|^{2} = \frac{1}{2\pi}
			\sum_{n,l}
			\frac{dp^{c}_{q_{e}}(nl \to E_{e})}{dE_{e}},
\end{align}
where $n,l$ are the quantum numbers of the emitted electron, $p^{c}_{q_{e}}(nl \to E_{e})$ estimates the probability to emit an electron with final kinetic energy $E_e$. Crucially  $q_e$ is related to the nuclear recoil energy $E_R$ through $q^2_e=2m_e^2 E_R/m_N$, capturing the $E_R$ dependence of the differential ionization probability $|Z_{\rm ion}|^{2}$. We have used the results from Ref.~\cite{Ibe:2017yqa} to estimate $\left|Z_{\rm ion}\right|^{2}$ for our analysis. In Eq.~\eqref{eq:ionrate} we have also included the detection efficiency  and the detector energy resolution is taken into account by normalized Gaussian smearing function~\cite{Aprile:2020yad, Aprile:2020tmw}, with standard deviation $\sigma(E_{\rm rec})/E_{\rm rec}= 0.3171/\sqrt{E_{\rm rec}[\rm keV]}+0.0015$. Expectedly the detection efficiency  is a function of the reconstructed electronic energy $E_{\rm rec},$ the effective energy observed at the detector. The ejection of electron from an inner orbital takes the resulting ion to an excited state. Subsequent de-excitation to the ground state is achieved by the release of electronic energy in the form of photons or additional electrons. Therefore the total energy deposited in the detector is approximately $E_{\rm EM} = E_{e} + E_{nl}$, with $E_{nl}$ being the binding energy of the electron before emission. Finally one can obtain the total rate by integrating Eq.~\eqref{eq:ionrate} within appropriate limits coming from energy-momentum conservation considerations. 
%
\begin{figure*}[!htbp]
\begin{center}
\subfloat[\label{sf:msigSI}]{\includegraphics[scale=0.20]{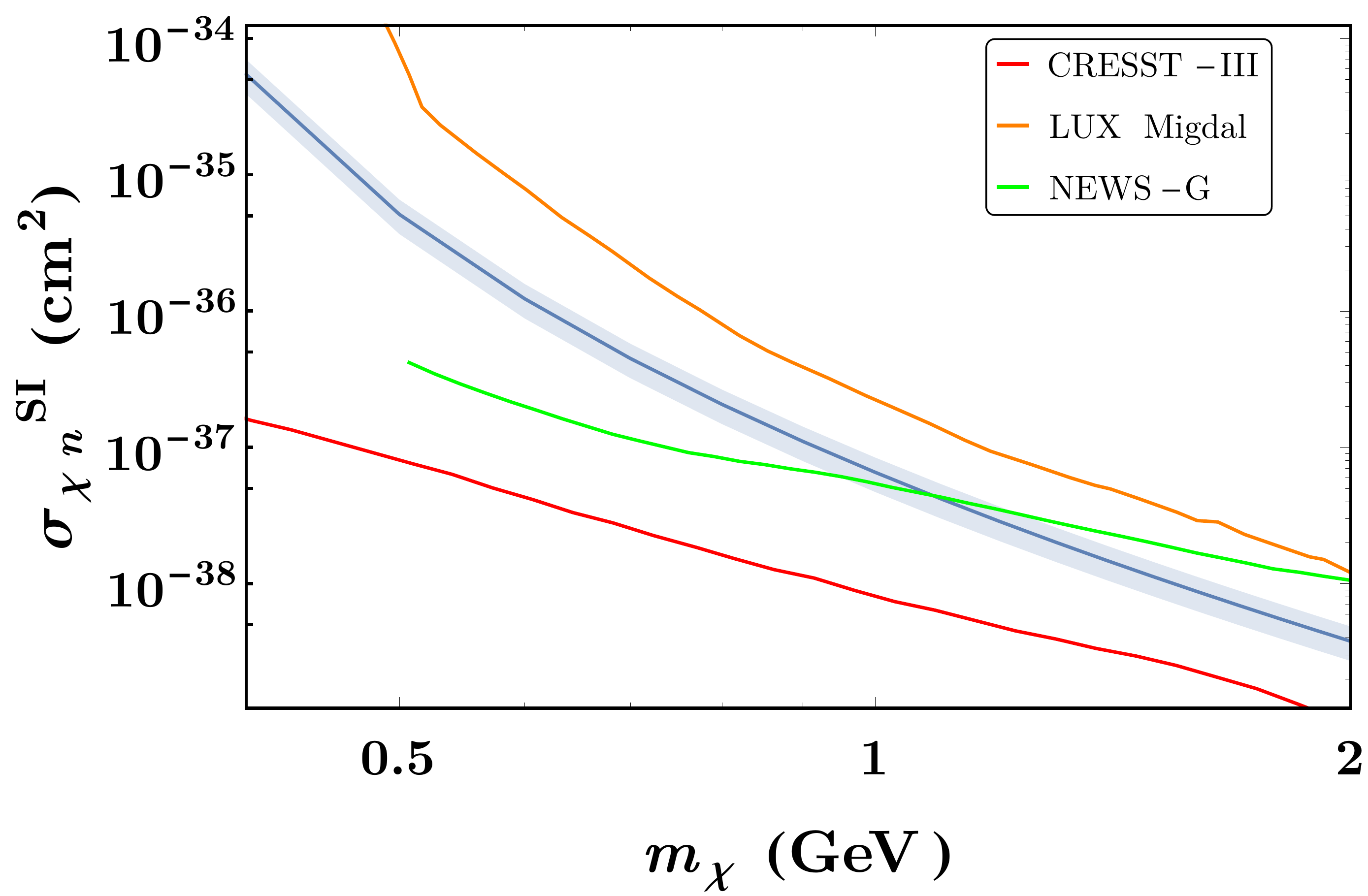}}~~~
\subfloat[\label{sf:msigSDp}]{\includegraphics[scale=0.20]{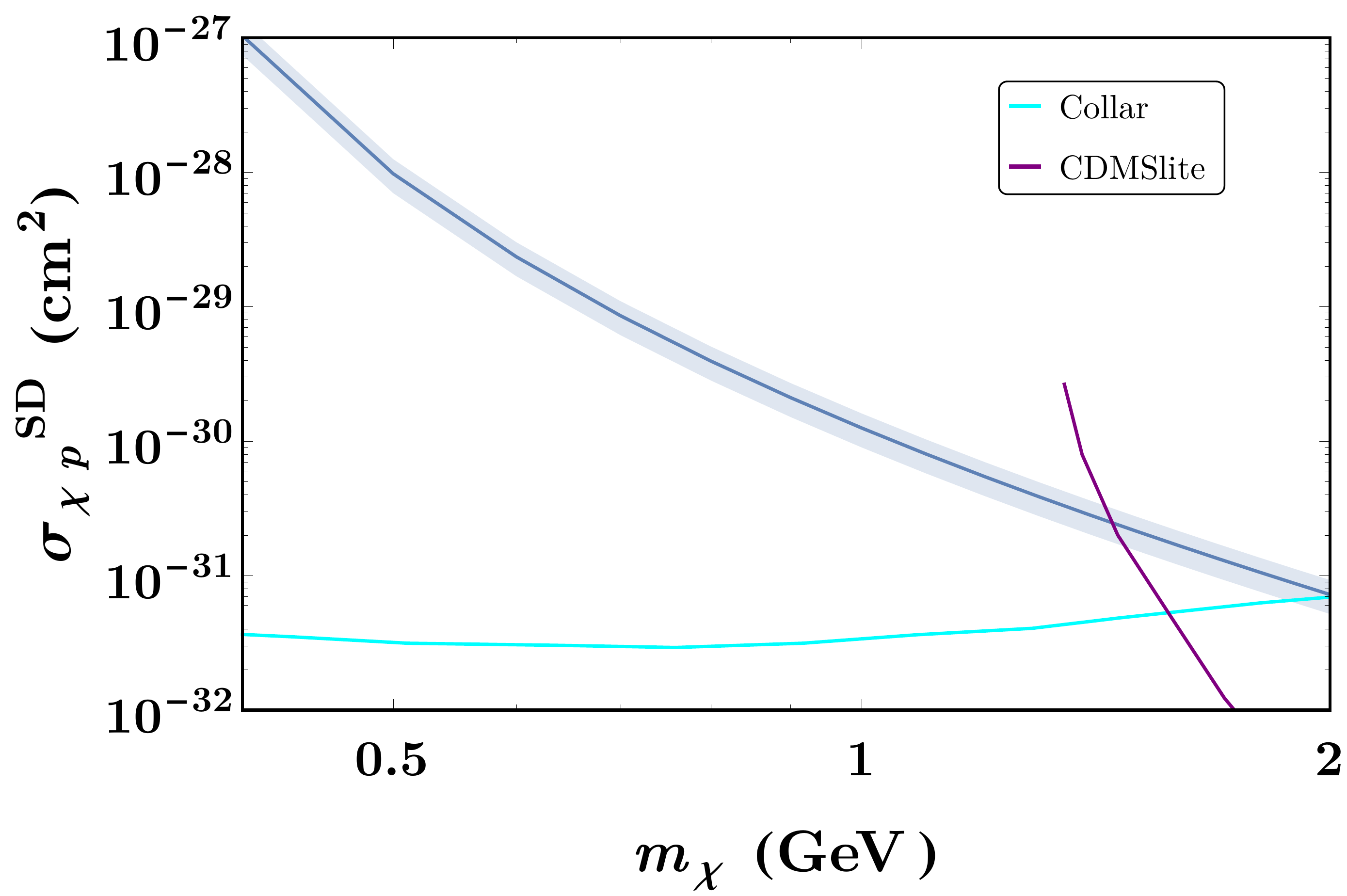}} 
\\
\subfloat[\label{sf:msigSDne}]{\includegraphics[scale=0.20]{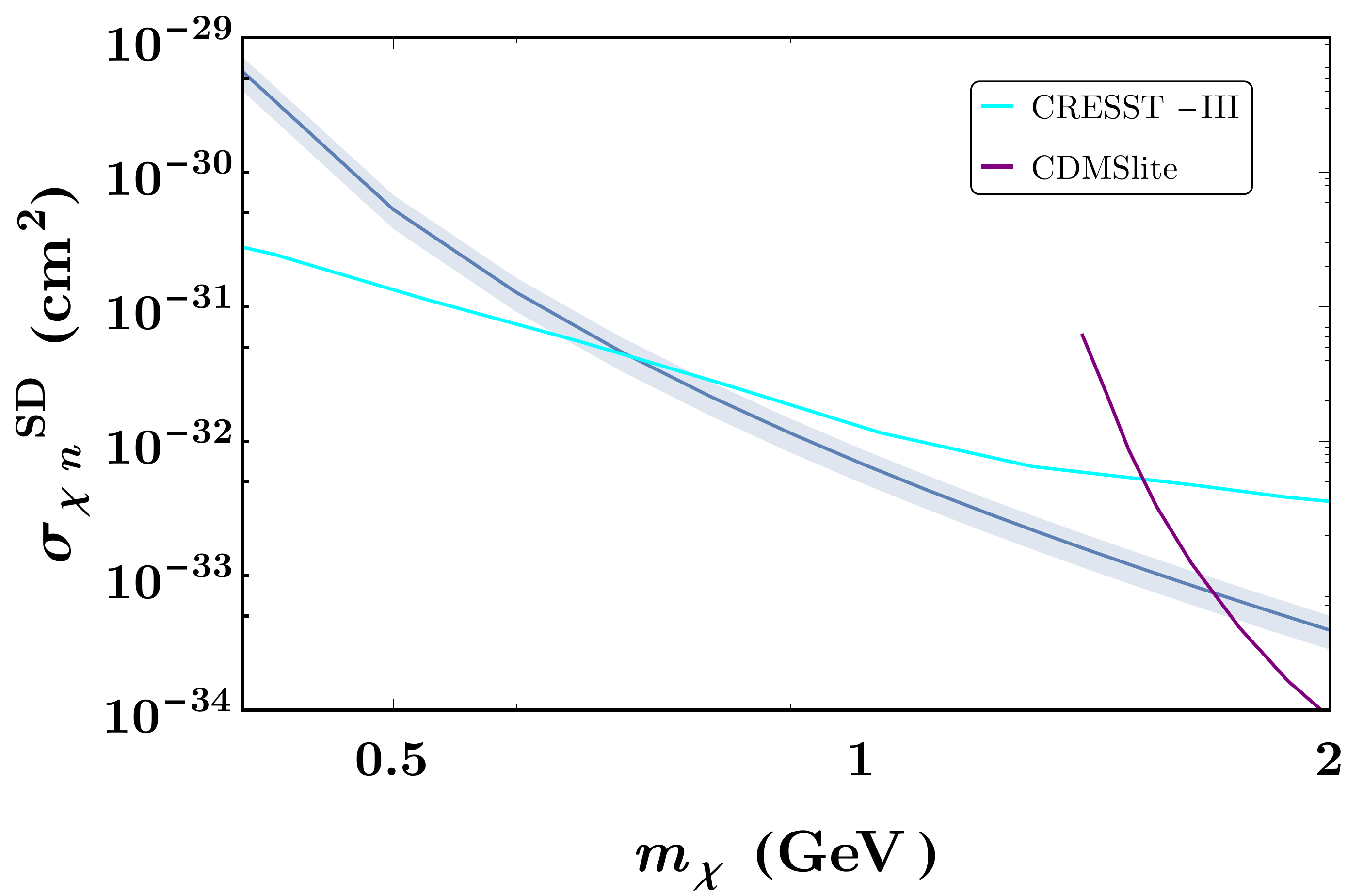}} ~~~
\subfloat[\label{sf:msigBDM}]{\includegraphics[scale=0.20]{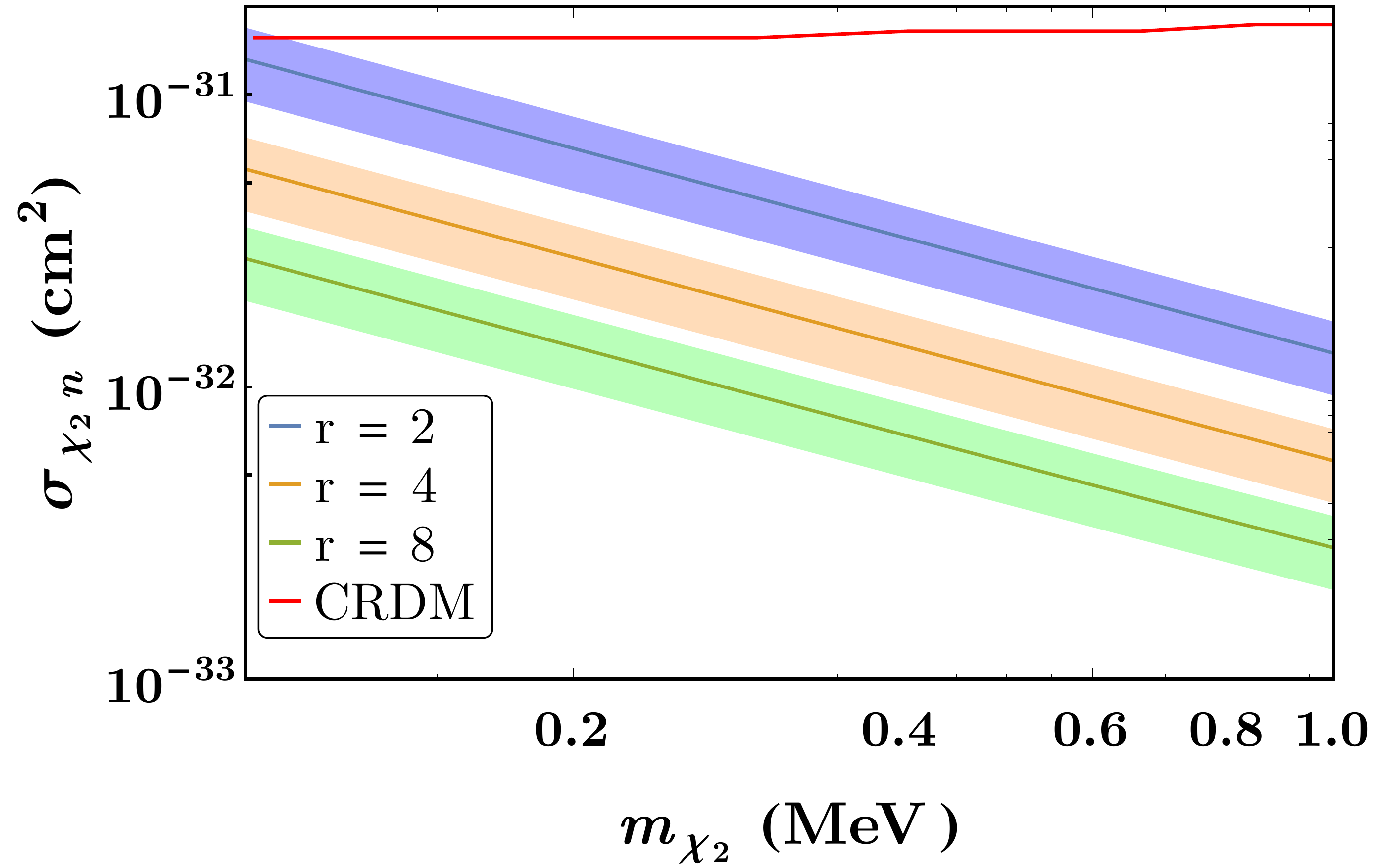}}
\caption{Constraints on parameter spaces in various cases. Vanilla DM scenario: the blue curve in (a), (b) and (c) is the contour consistent with 53 Migdal events in the 1-7 keV bin and the light-blue shaded region represents the $1~\sigma$ error in the background estimation of XENON1T. (a) Spin-independent case; the bounds from CRESST-III, LUX and NEWS-G searches are shown by red, orange and green curves respectively. (b) Spin-dependent (DM-proton): Constraints from anomalous cosmic ray search and CDMSlite are shown by cyan and purple curves respectively. (c) Spin-dependent (DM-neutron); CRESST-III and CDMSlite constraints are shown by cyan and purple curves respectively. (d) Boosted DM scenario: parameter space in $m_{\chi_{2}}$-$\sigma_{\chi_{2}n}$ plane yielding required number of events to explain the excess. The three bands correspond to three mass ratios $r = m_{\chi_{1}}/m_{\chi_{2}}$ = 2, 4, 8.  We also show the constraint on the cross-section coming from cosmic ray-dark matter scattering~\cite{Bringmann:2018cvk} by the red solid line.}
\label{fig:fig1}
\end{center}
\end{figure*} 
%
In passing we would like to comment on the possible principle quantum number $n$ relevant for our study. Since the available analysis of the Migdal effect is performed for the case of isolated atoms, the shifts in electronic energy levels due to atoms in a liquid is not taken into account. Given the energy scale of interest we explore the possibility of Migdal ionization from $n=3,4$ level. However, a possibility of utilising even the $n=2$ electrons in the case of boosted DM will be discussed later in this section. Note that all other energy levels do not contribute significantly in the region of observed signal and therefore has been neglected from our analysis.  
%

\subsection{Vanilla DM scenario}
\label{subsec:normDM}

We will now consider the  case of a  vanilla DM particle $\chi$ scattering with the Xe nucleus in the XENON1T detector giving rise to observed electron recoil via the Migdal effect. We take an model independent approach and consider the various velocity and momentum independent operators in the non-relativistic effective field theory (NREFT) of DM nucleon interactions, in turns \cite{Fitzpatrick:2012ix}. The DM mass $m_{\chi}$ and the spin (in)dependent scattering cross-section are considered as independent free parameters. For a given DM mass and cross section, we can now calculate the event rate by following the prescription laid out above.  We choose a range of  $m_{\chi}$ to keep the nuclear recoil energy below the XENON1T threshold while being able to ionize at least the $n=3,4$ electron from the Xe atoms. We set  the  nucleon-DM cross-section by matching the observed events using Eq.~\eqref{eq:ionrate}. 
The relevant parameter region for spin-independent interaction is presented in Fig.~\ref{sf:msigSI}. We find that the required cross-section is several orders of magnitude above the limit from CRESST-III experiment \cite{Abdelhameed:2019hmk} that sets the most stringent bound in this mass range. In addition to that we also show the bounds from NEWS-G~\cite{Arnaud:2017bjh} and LUX~\cite{Akerib:2018hck} searches.
Note that in this region of parameter space there exists an analysis of the XENON1T S2-only data~\cite{Aprile:2019jmx}. Since the same data set in the updated analysis~\cite{Aprile:2020tmw} shows the electron recoil excess in S1-S2 data, we do not impose the constraints of~\cite{Aprile:2019jmx} on the plots presented in Fig.~\ref{fig:fig1}.
We represent the SD parameters space for the neutron and proton case individually in Figs.~\ref{sf:msigSDp} and \ref{sf:msigSDne}. The relevant constraints from anomalous cosmic ray searches (Collar)~\cite{Collar:2018ydf} and low-mass DM searches with CDMSlite~\cite{Agnese:2017jvy} and CRESST-III are also shown. We find that for scenarios where DM-proton couplings are heavily suppressed \cite{Kelso:2017gib}, a DM mass between $0.72-1.67$ GeV is capable of explaining the excess observed at the XENON1T experiment while remaining consistent with all relevant constraints.

\subsection{Boosted DM Migdal scenario}
\label{subsec:boostDM}
In this section we discuss the possibility of Migdal events arising from boosted dark matter. These can excite the Migdal ionisation while remaining deep in the sub-GeV DM mass scale where the traditional DM direct detection  experiments do not have existing limits from standard nuclear-recoil or Migdal studies. Interestingly, the boost may be able to excite the $n=2$ ionisation with a peak at 4.9 keV, leading to a richer phenomenology.

In a  generic multi-component DM scenario with sufficient mass gap between multiple DM species, a minor component of DM (produced non-thermally by late-time decay of the heavier component) can be  boosted. These states having a non-thermal energy distribution can have sufficient flux to generate events at DM direct detection experiments.
We pare down this framework to consider a two-component DM scenario with heavier species $\chi_{1}$ and lighter $\chi_{2}$. To estimate the differential rate of events due to annihilation of $\chi_1$ to $\chi_{2}$ and subsequent  $\chi_{2}$-nucleus interaction in a direct detection experiment we closely follow~\cite{Cherry:2015oca},
\begin{align}
\frac{dR}{dE_{R}} = \frac{\Phi_{\chi_{2}}}{m_{N}}
            \int_{E_{\rm min}(E_{R})}^{\infty}
             dE_{\chi_{2}}\frac{dN}{dE_{\chi_{2}}}
             \left(\frac{d\sigma_{\chi_{2}N}}{dE_{R}}\right),
\end{align}
where $\Phi_{\chi_{2}}$ is the local flux of $\chi_{2}$-species, $E_{\rm min}(E_{R}) = \sqrt{m_{N}E_{R}/2}$ is the minimum energy required to produce a recoil of energy $E_{R}$, and $\frac{dN}{dE_{\chi_{2}}} = 2\delta(E_{\chi_{2}} - m_{\chi_{1}})$. The differential cross-section can be written as,
\begin{align}
\frac{d\sigma_{\chi_{2}N}}{dE_{R}} = \frac{G_{Y}^{2}}{2\pi}
		A^{2}m_{N}F^{2}(E_R)\left(1 - 
		 \left(
		   \frac{E_{\rm min}}{E_{\chi_{2}}}\right)^{2}
		 \right),
\end{align}
where $G_Y$ is the effective coupling, $A^{2}$ is the coefficient for the coherent enhancement of scattering with equal rates on neutrons and protons, and $F^{2}(E_R)$ is the nuclear form factor. We have taken the relevant flux for this kind of boosted DM from Ref.~\cite{Agashe:2014yua}.
In Fig.~\ref{sf:msigBDM} we present the parameter space consistent with the observed excess events.  The mass of the boosted DM species $\chi_{2}$, is varied in the MeV range for specific choices of the mass ratios ($r = m_{\chi_{1}}/m_{\chi_{2}}$) that essentially control the boost factor. In the plot each band represents the  1 $\sigma$ error in the number of events and each solid line for $r=$2, 4, 8 represents the parameters that can give rise to 53 events. The region above the red horizontal line is excluded by the constraint reported in~\cite{Bringmann:2018cvk}.
%
%
%
\begin{figure}[t]
\begin{center}
\includegraphics[scale=0.26]{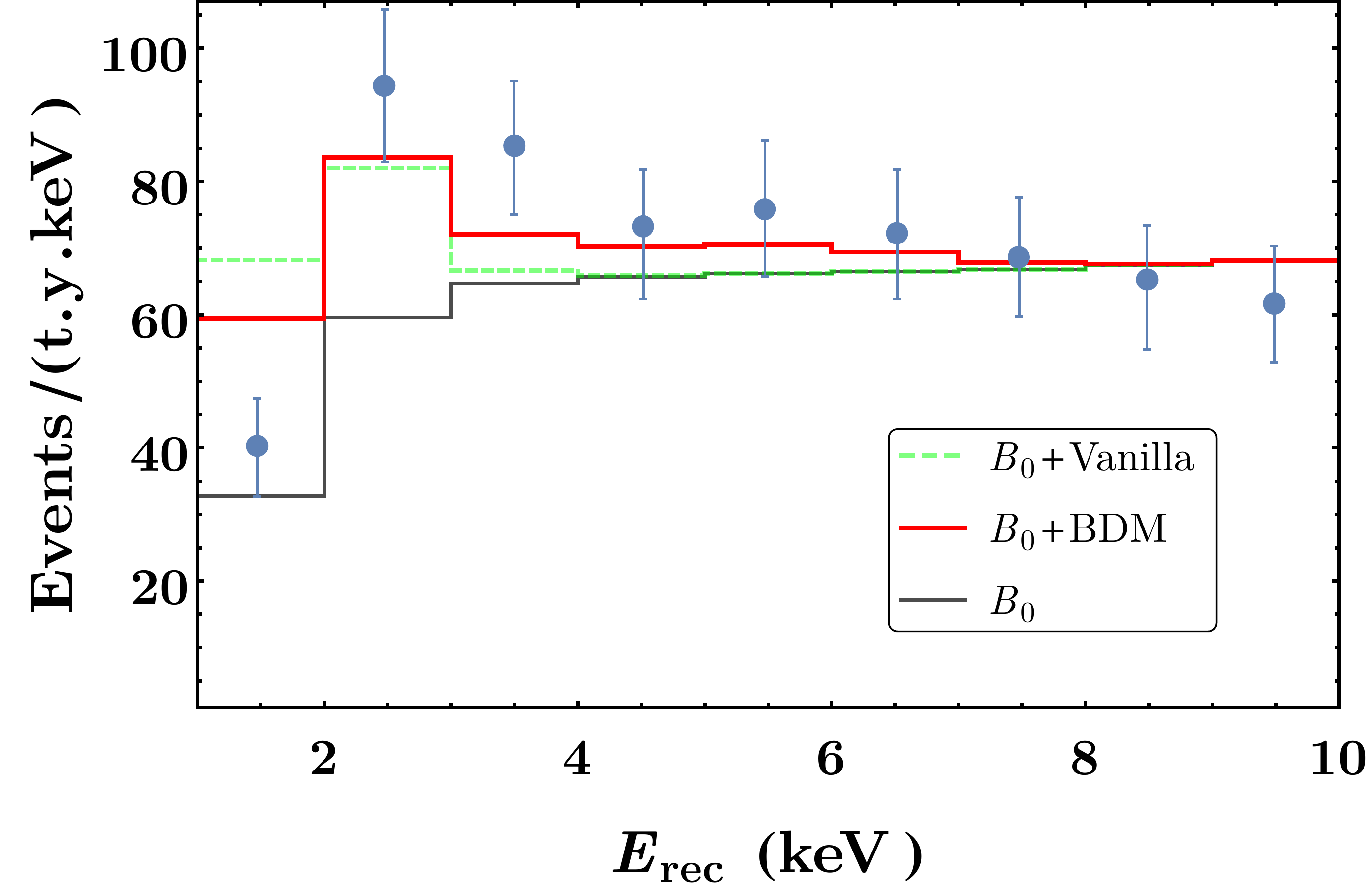}
\caption{Fitting the XENON1T data for the boosted (red solid) and vanilla (green dashed) DM scenarios. The binned background model $B_{0}$ is taken from~\cite{Aprile:2020tmw, Bloch:2020uzh}.}
\label{fig:evtRate}
\end{center}
\end{figure}

In Fig.~\ref{fig:evtRate} the red solid line represents a fit to the differential event rate data by taking an allowed benchmark point ($m_{\chi_1}=0.8 ~\text{MeV}, ~ r=8,~ \sigma_{\chi_{2} n}=1.1 \times 10^{-31} \text{cm}^2$) from this boosted scenario that reproduces the 53 observed events. Note that the fitting of the differential rate for boosted DM scenario is better compared to the vanilla scenario which is shown by the green dashed line in Fig. \ref{fig:evtRate}. This can be attributed to the additional contribution from the ionization of the $n=2$ electrons  in the boosted scenario. The Migdal contribution is expected to peak at  the threshold however, the sharp decline in the  detector efficiency in the first bin region  modulates the binned signal to take the form depicted in Fig.~\ref{fig:evtRate}. The binned fit for the boosted DM scenario is significantly better than the background-only hypothesis~\cite{Aprile:2020tmw, Bloch:2020uzh}. Admittedly, the solar axion or the neutrino magnetic moment explanation of the excess \cite{Aprile:2020tmw} yields a better fit owing to the fact that the  unreconstructed Migdal events peaks at the first bin.

\section{Conclusion} 
\label{sec:concl}
The very recent announcement of an excess  in the electron recoil events at XENON1T detector may be the first tantalising hint of the existence of particulate DM. While further data collection will clarify this situation, the unaccounted tritium background and proximity to the experiment threshold remains a cause for concern. In this letter we critically examine the possibility of Migdal effect  ionization events from DM-nucleon scattering, with below threshold nuclear recoils,  as a possible source for the observed events. It is demonstrated that a DM mass between  $0.72$ to $1.67$ GeV with a spin dependent neutron-philic interaction with the Xe nucleus can explain the observed excess events. An improved fit to the excess may be obtained where a boosted DM drive the Migdal effect. 

The Migdal excess are closely related to the atomic structure of the Xe atom, thus  any explanation of the present excess within this framework would entail the existence of additional correlated signal structures. Most prominent would be the events generated from the excitation of the $n=4$ atomic levels  electrons of the Xe atom with a peak at $61$ eV. Search for these peaks in present or future experiments may provide complementary evidence for the Migdal explanation of the XENON1T excess presented in the paper.

%

\paragraph*{Acknowledgements\,:} 
We thank A. Banerjee for the discussions. TNM thanks S. Dasgupta and R. Paramanik for assistance with the numerical calculations. TNM  acknowledges the Department of Science and Technology, Government of India, under the Grant Agreement No.  IFA13-PH-74 (INSPIRE Faculty Award) for a fellowship. TSR  is partially supported by the  Department  of  Science  and  Technology, Government of India, under the Grant Agreement No. ECR/2018/002192 (Early Career Research Award).


\bibliographystyle{JHEP}
\bibliography{xenonmigdal_ref.bib}
\end{document}